 \documentstyle[12pt]{article}
 
\def\eop{\vspace*{\fill}\pagebreak}

\newcommand{\newsection}{
\setcounter{equation}{0}
\section}
\def\tr{\,{\rm tr}\,}
\def\E{{\,\rm e}\,}
\def\be{\begin{equation}}
\def\ee{\end{equation}}
\def\bea{\begin{eqnarray}}
\def\eea{\end{eqnarray}}
\def\LA{\left\langle}
\def\RA{\right\rangle}
\newcommand{\rf}[1]{(\ref{#1})}
\def\a{\alpha}
\def\b{\beta}
\def\m{\mu}
\def\s{\theta}
\def\d{\partial}
\newcommand{\dd}[1]{\frac{\d}{\d{#1}}}
\newcommand{\DD}[2]{\frac{\d{#1}}{\d{#2}}}
\def\D{\delta}
\def\t{\tau}
\def\r{\rho}
\def\c{\sigma}

\begin{document}

\begin{titlepage}
\begin{flushright}
SMI-94-7 \\
May, 1994
\end{flushright}
\vspace{.5cm}

\begin{center}
{\LARGE Collective field approach }
\end{center}
\vspace{0.2cm}
\begin{center}
{\LARGE to gauged principal chiral field at large $N$}
\end{center}
\vspace{1cm}
\begin{center}
{\large K.\ Zarembo}
\footnote{E--mail:   \ zarembo@qft.mian.su \ }\\
 \mbox{} \\ {\it Steklov Mathematical Institute,} \\
{\it Vavilov st. 42, GSP-1, 117966 Moscow, RF}
\end{center}

\vskip 1 cm
\begin{abstract}
The lattice model of principal chiral field interacting with the gauge
fields, which have no kinetic term, is considered. This model can be
regarded as a strong coupling limit of lattice gauge theory at finite
temperature. The complete set of equations for collective field variables
is derived in the large $N$ limit and the phase structure of the model is
studied.
 \end{abstract}

\vspace{1cm}
\noindent

\eop
\end{titlepage}

\section{Introduction}

The Kazakov--Migdal model \cite{KM} was originally proposed as a model of
induced QCD. Although this idea suffers from several problems \cite{KSW},
there is another motivation for study of this model and
it's generalizations -- the usual for matrix models interpretation in terms
of random surfaces, $2d$ gravity and noncritical string theory. The
Kazakov--Migdal model significantly simplifies in the large $N$ limit and
can be treated by saddle point methods \cite{Mig,Gr}, or by means of loop
equations \cite{DMS}. In this paper the modification of the
Kazakov--Migdal model with the Hermitean matrix field replaced by the
principal chiral one is considered:
\be
{\cal
L}=N\tr\left[\frac{1}{2\gamma}\,(\nabla_{\m}g)^{\dagger}
(\nabla^{\m}g)-V(g)-V(g^{\dagger})\right],~~
\nabla_{\m}g=\d_{\m}g-[A_{\m},g].
\label{lagrangian}
\ee
There $g(x)$ is $U(N)$ (or $SU(N)$) matrix, while $A_{\m}(x)$ is $u(N)$
(or $su(N)$) gauge field.

This model, defined on a $D$-dimensional lattice, can be regarded as a
strong coupling limit of the $(D+1)$-dimensional lattice gauge theory at
finite temperature \cite{Pol,PSG,CDP}. It can be shown in a
most simple way within the Hamiltonian framework \cite{Pol}. In the
strong coupling limit only the electric part of the Kogut--Susskind
Hamiltonian \cite{KS}
\be
{\cal H}=-\frac{g^2}{2}\sum_{x,\m}\triangle_{U_{\m}(x)}-
\frac{2}{g^2}\sum_{\Box}\tr\left[U(\Box)+U^{\dagger}(\Box)\right]
\label{ks}
\ee
contributes to the partition function
\be
{\cal Z}={\rm Tr}\E^{-\frac{{\cal H}}{T}}.
\label{ftpart}
\ee
There $\triangle_{U_{\m}(x)}$ is invariant Laplacian acting on the group
variable $U_{\m}(x)$, attached to the link $(x,x+\m)$ of a $D$-dimensional
lattice. If it were not for the gauge invariance, the partition function
\rf{ftpart} would factorize in the strong coupling limit on the product
over the links of the heat kernels on the group manifold, independently
integrated over the gauge fields. However, one should first average over
all gauges:
\be
{\cal Z}_{s.c.}=\int\,\prod_{x,\m}DU_{\m}(x)\,\prod_x Dg(x)\,\prod_{x,\m}
K\left(U_{\m}(x),g^{\dagger}(x)U_{\m}(x)g(x+\m)\left|\frac{g^2N}{T}
\right.\right),
\label{1scpart}
\ee
where $K(g,h|\t)$ is a solution to
the heat equation
\be
\dd{\t}K(g,h|\t)=\frac{1}{2N}\triangle_{g}K(g,h|\t)
\label{heat}
\ee
with initial condition
\be
K(g,h|0)=\D(g,h).
\label{initial}
\ee
An explicit expression for $K(g,h|\t)$ in terms of a character expansion
is
\be
K(g,h|\t)=\sum_R\,{\rm dim}\,R\,\E^{-\frac{c_R\,\t}{2N}}
\chi_R(gh^{\dagger}),
\label{ch}
\ee
where the sum is taken over all irreducible representations of the gauge
group, with ${\rm dim}R$, $c_R$ and $\chi_R$ being the dimension,
quadratic Casimir and the character of the representation $R$.

The partition function \rf{1scpart} is nothing that the lattice
regularization of  \rf{lagrangian} with zero potential and $\gamma$
related to the gauge coupling $g$ and the temperature $T$ by
\be
\gamma=\frac{g^2N}{T}.
\label{gamma}
\ee
This model, with exponential Boltzmann weight instead of the heat kernel
one and a kinetic term for the gauge fields added, was studied both
numerically and in a mean field approximation for $D=3,4$ and
$N=2,3$ \cite{DJK}. The purpose of the present paper is to derive the
large $N$ equations of motion for collective variables in spirit of
Refs.\cite{JSD,JS} and to study the phase structure of the model in the
large $N$ limit.

It is well known that at some critical temperature lattice gauge theories
undergo the deconfining phase transition \cite{Pol,SY} associated with
the center group symmetry breaking. In terms of the above model this is
the symmetry with respect to the multiplication of all $g(x)$ by an
element of the center of the gauge group (this symmetry is spoiled by an
addition of the potential term). At low temperatures (large $\gamma$) the
model is in the symmetric (confining) phase and $\LA\frac1N\tr g(x)\RA$=0,
while at some critical value of $\gamma$ the symmetry is violated and the
chiral field gains a nonzero expectation value. It will be shown below,
that, at least for sufficiently large $D$, confining and deconfining
phases are separated by an intermediate one.

At high temperatures the
fluctuations of the chiral field are suppressed, as
 $\gamma$ is small, so it
can be expanded around unity
\be
g(x)=\E^{i\gamma\Phi(x)}\simeq 1+i\gamma\Phi(x)
\label{expansion}
\ee
and \rf{1scpart} reduces to the partition function for the Kazakov--Migdal
model with critical potential \cite{CDP}.

\newsection{Large $N$ equations of motion}

Consider the partition function \rf{1scpart} with a potential term added:
\bea
Z&=&\int\,\prod_x Dg(x)\,\prod_{x,\m}DU_{\m}(x)
\E^{-\sum_x\,N\tr\left[V(g(x))
+V(g^{\dagger}(x))\right]}\nonumber
\\
&\times&\prod_{x,\m}
K\left(\left.g(x),U_{\m}(x)g(x+\m)U^{\dagger}_{\m}(x)\right|\gamma
\right).
\label{scpart}
\eea
Integration over the gauge fields goes independently on each link, so,
fixing the diagonal gauge: $g(x)={\rm diag}(\E^{i\a_k(x)})$ and
integrating out the link variables, one is left with an effective action
for the eigenvalues of $g(x)$, which contains only local interactions:
\bea
S_{eff}&=&\sum_x\left[N\sum_i U\left(\a_i(x)\right)-
\sum_{i<j}\log\sin^2\frac{\a_i(x)-\a_j(x)}{2}\right.\nonumber\\
&-&\left.\frac12\sum_{\m=-D}^D
\log I\left(\a(x),\a(x+\m)|\gamma\right)\right],
\label{seff}
\eea
where
\be
U(\s)=V(\E^{i\s})+V(\E^{-i\s}),
\label{u}
\ee
the second term in \rf{seff} comes from the decomposition of invariant
measure:
\be
Dg=D\Omega\prod_{i=1}^{N}d\a_i\,J^2(\a),~~J(\a)=\prod_{i<j}
\sin\frac{\a_i-\a_j}{2}
\label{jac}
\ee
for $g=\Omega \E^{i\a}\Omega^{\dagger}$, and
\be
I(\a,\a'|\t)=\int\,
DU\,K(\E^{i\a},U\E^{i\a'}U^{\dagger}\left|\t
\right.).
\label{int}
\ee

It is convenient to introduce the density of the eigenvalues of the matrix
$g(x)$:
\be
\r(\s,x)=\frac1N\sum_{i=1}^N \D\left(\s-\a_i(x)\right).
\label{dens}
\ee
In the large $N$ limit $\r(\s,x)$ becomes smooth $2\pi$--periodic
function, normalized to unity on the interval $(-\pi,\pi)$. For simplicity
we restrict our consideration to the case of $U(N)$ gauge group. For
$SU(N)$ the constraint $\int_{-\pi}^{\pi}\,d\s\,\s\r(\s,x)=0$ should be
imposed on the eigenvalue density and the Lagrange multiplier ensuring
this condition should be introduced. This is achieved by adding the term
$\lambda(x)\s$ to the potential \rf{u}.

The classical equations of motion for \rf{seff} are exact in the large $N$
limit and can be written in terms of the eigenvalue density as follows:
\be
U'(\s)+(D-1)G(\s,x)=\sum_{\m=-D}^{D}F_{\m}(\s,x|\gamma),
\label{eqm}
\ee
where
\be
G(\s,x)=\wp\int_{-\pi}^{\pi}\,d\s'\r(\s',x)\cot\frac{\s-\s'}{2}
\label{defg}
\ee
and we have defined
\be
F_{\m}(\s,x|\gamma)=\frac{1}{N^2}\dd{\s}\frac{\D}{\D\r(\s,x)}
\log I\left(\a(x),\a(x+\m)\left|\gamma\right.\right)+\frac12 G(\s,x).
\label{defv}
\ee

The one--link integral \rf{int} is the unitary analog of the
Itzykson--Zuber integral and can be treated by the same method, proposed
in Ref. \cite{IZ} for finite $N$ and used to calculate the large $N$
limit of the Itzykson--Zuber integral in Ref. \cite{Mat}. This method
exploits the relation of the matrix quantum mechanics to that of the free
nonrelativistic fermions \cite{BIPZ}. Really, $K(g,h|\t)$ is an imaginary
time transition function for the free Hamiltonian on the $U(N)$ group
manifold. After the calculations, similar to that of Ref. \cite{IZ}, it
can be shown, that $J(\a)J(\a')I(\a,\a'|\t)$ is the transition function
for $N$ noninteracting fermions on a circle. From which it immediately
follows, that it can be written in a form of a Slater determinant:
\be
I(\a,\a'|\t)=const\cdot\frac{\det_{kj}\vartheta\left(
\frac{\a_k-\a_j'}{2\pi}\left|\frac{i\t}{2\pi N}\right.\right)}
{J(\a)J(\a')}\, ,
\label{finite}
\ee
where $\vartheta(z,\t)$ is the Riemann theta function.

In the large $N$ limit the fermions behave semiclassicaly, so the
transition function for them is dominated by a single classical trajectory
given, in terms of the density $\c(\s,\t)$, by Euler equations for an
ideal fluid \cite{JSD,JS}:
\be
\DD{\c}{\t}+\dd{\s}(\c s)=0,
\label{cont}
\ee
\be
\DD{s}{\t}+s\DD{s}{\s}-\pi^2\c\DD{\c}{\s}=0.
\label{euler}
\ee
The minus sign before the last term in \rf{euler} is because the time is
imaginary. The function $s(\s,\t)$ is a velocity of the Fermi fluid. Eqs.
\rf{cont}, \rf{euler}, of course, can be derived directly by a  reduction
of the heat equation for \rf{int} to the Hamilton--Jacoby equation for the
collective field Hamiltonian by the calculations, analogous to that of
Ref. \cite{Mat}.

To find \rf{defv} one should solve \rf{cont}, \rf{euler} with
boundary conditions
\be
\c(\s,0;x,\m)=\r(\s,x),
\label{init}
\ee
\be
\c(\s,\gamma;x,\m)=\r(\s,x+\m).
\label{final}
\ee
Then for \rf{defv} one has:
\be
F_{\m}(\s,x|\gamma)=s(\s,0;x,\m).
\label{v}
\ee
This formula follows from the definition \rf{defv}, because the variation
of the classical action with respect to the initial condition for the
density \rf{init} is equal to it's conjugate variable, which is a
potential of the velocity \cite{LL}. Reversing time in \rf{cont},
\rf{euler} one also finds:
\be
F_{-\m}(\s,x|\gamma)=-s(\s,\gamma;x-\m,\m).
\label{v-}
\ee

Thus in the large $N$ limit the model reduces to the set of differential
equations \rf{cont}, \rf{euler} to be solved on each link with boundary
conditions for the density given by \rf{init}, \rf{final} and the sum of
the jumps of the velocities at each lattice cite given by the l.h.s. of
eq. \rf{eqm}. Eqs. \rf{cont}, \rf{euler} are integrable, although in an
implicit form, and the problem is, in principle, reducible to the purely
algebraic one \cite{Mat}, differing from that for Hermitean matrix model
in the periodic boundary conditions for $\s$ and the kernel in the
integral \rf{defg} (the Cauchy one versus the Hilbert kernel in the
Hermitean case).

In one--dimensional case the gauge field in \rf{lagrangian} can be
absorbed by a gauge transformation, so $D=1$ model is equivalent to the
unitary matrix quantum mechanics, solved in the large $N$ limit in Refs.
\cite{JS,Wadia}. It is instructive to reproduce these results from the
above lattice equations. One can suppose that $\t$ in eqs. \rf{cont},
\rf{euler} changes from $-\infty$ to $+\infty$, while at the points
$\t=n\gamma$ the velocity has a jump according to \rf{eqm},
\rf{v}, \rf{v-}.
However, in the continuum
limit $\gamma$ and $U'(\s)$ scales as $a\gamma$ and
$aU'(\s)$, respectively ($a$ is the lattice spacing), so the
discontinuities in $s(\s,\t)$ can be absorbed by adding the term
$\frac{1}{\gamma}U'(\s)$ to
the r.h.s. of eq. \rf{euler}. The magnitude of a
deviation of the solution of resulting equations from the exact one on
the time interval
$\left(na\gamma,(n+1)a\gamma\right)$ is of the order $(a\gamma)^2$
and can be neglected as $a\rightarrow 0$. Thus one obtains in the
continuum limit:
\be
\DD{\r}{t}+\gamma\dd{\s}(\r v)=0,
\label{contc}
\ee
\be
\DD{v}{t}+\gamma v\DD{v}{\s}+\pi^2\gamma\r\DD{\r}{\s}+U'=0,
\label{eulerc}
\ee
where $t$ is the real time, measured in an ordinary units (while $\t$ is
measured in the units of $\gamma$).
These are nothing that Jevicki--Sakita
collective field theory equations \cite{JS}.

In the multidimensional case the second term in the l.h.s. of eq. \rf{eqm}
do not vanish and is of order unity as $a\rightarrow 0$, so to obtain the
continuum limit in the same way as in one dimension, the solutions should
be considered with the velocity $s(\s,\t;x,\m)$ being by itself of order
$a^{-1}$ and $\gamma$ scaling
as $a^2\gamma$. Then the arguments, similar to that
used to obtain \rf{contc}, \rf{eulerc}, give the following equations in
the continuum limit (in the Minkowski space):
\be
\DD{\r}{x^{\m}}+\gamma\dd{\s}(\r v_{\m})=0,
\label{contmd}
\ee
\be
\DD{v^{\m}}{x^{\m}}+\gamma v_{\m}\DD{v^{\m}}{\s}+(D-1)G+U'=0.
\label{eulermd}
\ee
As in the case of Hermitean matrix model, it can be shown, that
translationally invariant solution to these equations is unstable
\cite{Zar}, the spectrum containing an infinite number of tachyons. Thus
one has to look for other possibilities to take the continuum limit. In
this respect the study of the phase structure of the model is of
fundamental importance.

\newsection{Strong coupling phase}

Consider the model \rf{scpart} without a potential term. Equations of the
previous section always have a trivial solution $\r(\s,x)=\frac{1}{2\pi}$
in this case (with $s=0$ in \rf{cont}, \rf{euler}). This solution is valid
at large $\gamma$. However, the instability
in the continuum limit at $\gamma=0$
shows, that at some
$\gamma_c$ the strong coupling solution becomes a maximum,
rather than a minimum, of the effective action.

To find the spectrum of excitations about the strong coupling solution we
write
\be
\r(\s,x)=\frac{1}{2\pi}+\frac{1}{2\pi}\sum_{n\neq 0}a_n(x)\E^{in\s},~~
a_n^*=a_{-n}
\label{ldens}
\ee
\be
\c(\s,\t;x,\m)=\frac{1}{2\pi}+\frac{1}{2\pi}\sum_{n\neq
0}\a_n(\t;x,\m)\E^{in\s},~~ \a_n^*=\a_{-n}
\label{lsigma}
\ee
\be
s(\s,\t;x,\m)=\sum_{n\neq
0}\b_n(\t;x,\m)\E^{in\s},~~\b_n^*=\b_{-n}
\label{ls}
\ee
and linearize the equations of motion in $a_n$, $\a_n$ and $\b_n$. The
solution  to     eqs.     \rf{cont},     \rf{euler}     then      reads:
\be
\a_n(\t;x,\m)=\a_n^+(x,\m)
\E^{\frac{n\t}{2}}+\a_n^-(x,\m)\E^{-\frac{n\t}{2}},
\label{alpha}
\ee
\be
\b_n(\t;x,\m)=
\frac{i}{2}
\left[\a_n^+(x,\m)\E^{\frac{n\t}{2}}-\a_n^-(x,\m)\E^{-\frac{n\t}{2}}
\right].
\label{beta}
\ee
Substituting  these  expressions  in  the  boundary  conditions  \rf{init},
\rf{final}, \rf{v}, \rf{v-} and \rf{eqm}  one obtains, after a  simple
algebra, the equation  for  the   Fourier  coefficients  of   the
eigenvalue density:
\be
\sum_{\m=1}^{D}\left[a_n(x+\m)-2\left(\cosh\frac{n\gamma}{2}-
\frac{D-1}{D}\sinh\frac{n\gamma}{2}\right)a_n(x)+a_n(x-\m)\right]=0,
\label{laplas}
\ee
which      leads      to       the      following      mass       spectrum:
\be
M_{n}^{2}=D\cosh\frac{n\gamma}{2}-(D-1)\sinh\frac{n\gamma}{2},
\label{mass}
\ee
where  $M_n^2=\sum_{\m=1}^{D}\cosh\lambda_{n}^{\m}$  for  the  solution  of
\rf{laplas}                  of                  the                   form
$\exp\left(\sum_{\m=1}^{D}\lambda_{n}^{\m}x^{\m}\right)$.

At the critical point
\be
\gamma_c=2\log(2D-1)
\label{gc}
\ee
the lowest excitation becomes massless, and the strong coupling solution
becomes unstable. In principle this critical point can be used to
construct the continuum limit. However, it describes only one complex
scalar particle with the mass
\be
m^2=(D-1)\frac{\gamma-\gamma_c}{a^2},
\label{cmass}
\ee
while the masses of other excitations remain of the cut--off order. It is
worth mentioning, that
an obstacle to take the continuum limit at $\gamma_c$
may occur due to the possibility for the model to undergo a first order
phase transition into the weak coupling phase before the point \rf{gc} is
reached (see the next section).

\newsection{Weak coupling phase}

As $\gamma\rightarrow 0$
the expansion \rf{expansion} becomes legitimate, so
one expects that the eigenvalue density $\r(\s,x)$ is peaked about zero.
Thus, in the zero approximation, the cotangent in \rf{defg} can be
replaced by $\left(\frac{\s-\s'}{2}\right)^{-1}$ and the equations of
motion become the same as for the Kazakov--Migdal model with the action
containing only the derivative squared term. It is known, that such model
has a stable translationally invariant solution \cite{Gr}. The
fluctuations about it contains one massless mode \cite{AG} related to the
shifts of the center of the eigenvalue density. It is straightforward to
show (in the general case) that, if $\r(\s,x)$ solves the equations of
Sec. 2, then $\r\left(\s+\s_0(x),x\right)$ also does, provided that
$\s_0(x)$ satisfies lattice Laplace equation. So the Goldstone
mode simply decouples and can be disregarded (one may exclude it
considering $SU(N)$ model).

Now let us take into account the next term in the expansion of the
cotangent in \rf{defg} (the eigenvalue density is assumed to be an even
function of $\s$):
\be
\wp\int\,d\s'\r(\s',x)\cot\frac{\s-\s'}{2}=
2\,\wp\int\,d\s'\frac{\r(\s',x)}{\s-\s'}-\frac16\s-\ldots
\label{cot}
\ee
After the substitution of \rf{cot} in the l.h.s. of eq. \rf{eqm} the
equations of motion become equivalent to that for the Kazakov--Migdal
model with a mass term added. Note, that this mass is negative. Moreover,
the residual term in \rf{cot} is given by the convergent, as
$|\s-\s'|<2\pi$, series in the odd powers of $\s-\s'$ with negative
coefficients. Thus an account of the corrections coming from the
nonlinearity of the chiral field leads to the $\r$-dependent upside--down
potential and \rf{cot} gives an upper bound for it. For sufficiently weak
potential an interaction with the gauge fields  stabilizes the
fluctuations and provides a minimum for an effective action, but at some
critical $\gamma_*$ the effective potential becomes strong enough to make
a weak coupling solution unstable. An upper estimate on $\gamma_*$ can be
obtained considering the Kazakov--Migdal model with the quadratic
potential, an effective mass being given by (in the notations of Ref.
\cite{Gr}):  \be m_{eff}^2=2D-\frac16(D-1)\gamma, \label{meff} \ee the
solution to which is known \cite{Gr}:
\be
\r(\s)\simeq
\frac{1}{\pi}\sqrt{\m-\frac14\m^2\s^2}, \label{reff} \ee \be
\m=\frac{m_{eff}^2(D-1)+D\sqrt{m_{eff}^4-4(2D-1)}}{(2D-1)\gamma}\, .
\label{mueff}
\ee
This solution is stable until $m_{eff}^2=2\sqrt{2D-1}$, when \rf{mueff}
becomes complex and the massless excitation appears in the spectrum
\cite{AG}. This gives the following bound on $\gamma_*$:
\be
\gamma_*<\frac{12(D-\sqrt{2D-1})}{D-1}\, .
\label{bound}
\ee
It is worth mentioning that for $D=2,3$ the width of the support of
eigenvalue density \rf{reff} becomes larger than $2\pi$ for smaller value
of the coupling.

There are, in principle, three possibilities for  the critical behavior:
\begin{description} \item[i)]
$\gamma_*>\gamma_c$. In this case strong and
weak coupling solutions can coexist and the first order phase transition
takes place at some $\gamma$ between $\gamma_c$ and $\gamma_*$.
\item[ii)] $\gamma_*=\gamma_c$. The phase transition is of the second
order.  \item[iii)] $\gamma_*<\gamma_c$. This means that an intermediate
phase exists, separated from strong and weak coupling ones by the second
order phase transitions.  \end{description} The third possibility is
realized in the large $D$ limit. Really, $\gamma_c$ grows as $2\log D$,
while $\gamma_*$ tends to $12$ as $D\rightarrow \infty$ (the estimate
\rf{bound} gives
an exact large $D$ asymptotics of $\gamma_*$, because the
width of the eigenvalue distribution \rf{reff} behaves at the critical
point as $4\left(\frac{72}{D}\right)^{1/4}$ and corrections to \rf{cot}
vanish). However, using an upper bound \rf{bound} it is possible to
advocate the existence of an intermediate phase only for $D>86$. It is
worth mentioning that the value of $\s$, at which the eigenvalue density
\rf{reff} with critical $m_{eff}^{2}$ and $\gamma$,  given by an upper
bound \rf{bound}, turns to zero, is, nevertheless, not small --
$\s_{max}=0.66\pi$ for $D=87$, so the approximation \rf{cot} remains rough
 at the point of the phase transition even for such large $D$. Thus more
accurate treatment is necessary to distinguish between the possibilities
i)--iii) for low dimensions.

For $D=3,4$ the first possibility is favored by the Monte Carlo
simulations of Ref. \cite{DJK} and the large amount of numerical data on
the finite temperature gauge theories. However, the second possibility do
not contradict the numerical results, as we are dealing with the boundary
of the phase diagram in $g^2$, $T$, and the line of the first order
transitions may terminate at the point of the second order one. The third
possibility is also not excluded, because an intermediate phase can
disappear at some finite, but large $N$, and thus be invisible in the
Monte Carlo simulations for $N=2,3$.

\newsection{Conclusions}

If we discard the possibility of the first order phase transition between
the strong and the weak coupling phases, which is legitimate, at least
for large $D$, as discussed in Sec. 4, the continuum limit of the model
can be constructed.  Unfortunately, corresponding continuum theory
contains  only finite number of degrees of freedom, if we approach the
critical point from strong or from weak coupling phase, and thus cannot
describe extended objects, which was one of the motivations for study of
the model.  However, this may not be the case for the model with a
potential term added, or if one takes the continuum limit in the
intermediate phase. It is also interesting to understand the origin of
this phase from the point of view of thermal lattice gauge theory.  To
clarify these questions it is desirable to know an exact solutions to the
model in the weak and the intermediate phases. To this end, it would be
interesting to develop an alternative approach, based on Schwinger--Dyson
equations, which proved to be useful in the context of the Kazakov--Migdal
model \cite{DMS}.

The author is grateful to Yu. Makeenko for discussions. The work was
supported in part by RFFR grant No. 94-01-00285.

\newsection*{Note added}

When this work was being prepared for publication, there appeared a paper
by D.V.~Boulatov \cite{Boul}, which also includes the treatment of the
gauged chiral field model at large $N$.

\end{document}